# Exceptional Point Singularities in Multi-Section DFB Lasers

Mehran Shahmohammadi,[a)] Martin J. Süess, Romain Peretti, Filippos Kapsalidis, Andres Forrer, Mattias Beck, and Jérôme Faist

*Institute for Quantum Electronics, ETH Zürich, 8093 Zürich, Switzerland*

(Dated: 7 March 2022)

A laser exhibits both controllable gain and loss and, under proper design conditions, is an ideal non-Hermitian system allowing the direct observation and engineering of spectral singularities such as exceptional points (EPs). A dual section distributed feedback (DFB) quantum cascade laser (QCL) is a prototype of such a system, allowing the controlled coupling of a ladder of cavity Fabry-Perot (FP) modes to a quarter wave shifted DFB mode. Tuning the coupling strength and the gain difference between these two set of modes enables probing the regimes from weak coupling to strong coupling and the robust observation of exceptional point singularities. At these exceptional points, the laser exhibits a sequence of lasing and coherent prefect absorption dynamics[1,2] when pumped above transparency. Additionally, the pumping scheme allows the deliberate lifting of the exceptional point degeneracy. These results show that dual section QCL is a perfect platform to study exceptional points because the coupling parameter and system loss can be tuned in a single device.

[a)]correspounding author: shahmohammadimehran@gmail.com

Lasers are prototype non-Hermitian systems because they feature intrinsically gain and losses. The cavity modes within them can interact and couple to each other, and tuning of the gain/losses of supported modes by the laser cavity subsequently give rise to the occurrence of exceptional points (EPs), at which two or more eigenvalues and the corresponding eigenvectors coalesce. In contrast to the degeneracy in a Hermitian system (a diabolic point), an EP is a branch point in the complex energy surface where the Hamiltonian matrix becomes defective. EPs are of great importance due to the counter-intuitive phenomena associated with them, such as loss-induced optical transparency[3], unidirectional transport of light[3–6], topological energy transfer[7].

The importance of the EPs arises also from the fact that these singularities mark the boundary between the weakly and strongly coupled regime[8]. In coupled laser cavities, an exceptional point can lead to pump-induced[9] or loss- induced[10] revival of lasing. Also this phenomena was used for mode control of lasing in coupled micro-rings under asymmetric optical pumping[11]. Several systems have been explored in order to enable the efficient access of EPs such as coupled photonic crystal nanolasers, where employing asymmetric optical gains/losses leads to a phase transition of lasing modes at EPs[12]. Coupled micrometer-sized cavities was another platform in order to investigate EPs, where in addition to asymmetric gain or loss, one can also control the intercavity coupling strength[4,11]. Still a simplified and practical approach to access the EPs and investigate the phase transitions around this spectral singularity remained a challenge.

We have previously realized dual section DFB QCLs with electrically wavelength-switching capability[13] for the purpose of gas spectroscopy applications at mid-IR spectrum[14]. In our realized devices, two DFB grating sections featuring quarter wave shift (QWS) defect, each optimized on different target lasing wavelength, are stacked on top of a single ridge waveguide. In order to allow each section to be independently pumped, the top layer cladding was etched partially between the two segments and the top metal pads were realized separated (see Fig. 1(a)). Despite the great features of these devices for gas spectroscopy applications, they suffer from the distinct power modulations above threshold. The depth of these power modulations was so high in some cases, that could even switch on and off the device above the first lasing threshold. This strong power modulation was attributed to the etaloning effect, i.e. the interference of the light originated from the active DFB section inside the passive section. As shown by the finite element simulations in 1(b,c), the temperature profile of the laser can lead to relatively large difference temperature between the front and back section of the device. As a results of this temperature profile driven by the injected current, the passive section of laser acts as an etalon that its transmission spectrum is tuned with respect to the DFB mode, and therefore modulates the output power the laser. Although, this picture can well explain the case of weak coupling between the FP modes and the lasing DFB modes, it does not capture all dynamics of the laser as will be discussed in the following. Applying low reflectively anti-reflective (AR) coatings was consider as a remedy to suppress the modulations originated from etaloning effect of the passive section on the lasing mode power. Here, we demonstrate that the different coupling regimes of the FP modes and the QWS DFB mode can be interpreted through a non-Hermitian dynamics model featuring EPs for certain values of the parameters.

FP modes with an energy close to the one the target lasing mode experience the high reflection of the grating on one side and the cleaved laser facet reflection on other side, as shown as reflectors $r_2$ and $r_1$ in Fig.1(d)). Therefore, the intensity profile of the FP modes are localized at the passive section of device. On the other hand, the intensity profile of QWS lasing mode is localized at the QWS defect position in the grating. This separation of profile, allows careful engineering of the mode overlaps and therefore the coupling of these two set of modes. The design parameters, such as the relative position of the QWS position in the grating and its distance from the passive section, combined with the strength of the grating can be used tune the overlap of these modes. Moreover, the electrical injection in different segments allows tuning of the temperature profile and therefor the refractive index profile along the laser. As a direct consequence, the refractive index of the DFB grating at the interface of the two DFB segments can be distorted, and hence the FP modes confinements can be relaxed. This helps increasing the wavefunction overlap of FP modes with the QWS DFB mode, and allows electrical tuning of the coupling strength $\Omega$ of these two set of modes.

In order to make it easier to grasp the dynamics of the system, let us first look at the dynamics of the eigenvalues of a basic characteristic matrix (Hamiltonian) of a system, where one resonant mode interacts with a group of orthogonal modes. Typically, the coupled cavity systems are designed to be reduced to two dimensions, where only two modes are coupling with each other. However, the physics can be generalized and expanded to the case where more than two modes couples to each other in laser cavity. The non-Hermitian Hamiltonian of a system consisting of two coupled modes can be expressed as:

$$H = \begin{bmatrix} \omega_1^0 + ig_{net1} & \frac{\Omega}{2} \\ \frac{\Omega}{2} & \omega_2^0 + ig_{net2} \end{bmatrix}, \quad (1)$$

where $\omega_j^0$ denotes the angular frequency of the mode $j$ in absence of any coupling, and $g_{netj} = g_j - \alpha_{mj} - \alpha_{wj}$ denotes the net gains of each mode including the gain $g_j$, mirror losses $\alpha_{mj}$ and waveguide losses $\alpha_{wj}$. The coupling strength of the modes is given at the off-diagonal elements of the matrix by coupling constant $\Omega/2$. The eigenvalues of this system are described as:

$$\omega_{1,2} = \omega_{av.} + ig_{av.} \pm \sqrt{(\frac{\Omega}{2})^2 + (\Delta\omega - i\Delta g)^2} \quad (2)$$

, with $\omega_{av.} = \frac{\omega_1+\omega_2}{2}$ and $g_{av.} = \frac{(g_{net1}+g_{net2})}{2}$ standing for the mean values of resonance angular frequencies and the average net gains of the modes. The terms $\Delta\omega = \frac{(\omega_1-\omega_2)}{2}$ and $\Delta g = \frac{(g_{net1}-g_{net2})}{2}$ are the differences between the angular frequencies and the net gain of modes.

When the modes are tuned, i.e., $\Delta\omega=0$, the ratio $\frac{\Delta g}{\Omega}$ determine the system dynamics. This is is shown for the expanded version of Hamiltonian (1), where multiple FP modes coupled to a the modes originated from a DFB grating. Here, we simplified the modes DFB to a QWS defect mode of the DFB grating. The expanded non-Hermitian Hamiltonian reads:

$$H = \begin{bmatrix} \omega_{QWS} + ig_{QWS} & \frac{\Omega}{2} & \frac{\Omega}{2} & \frac{\Omega}{2} & \cdots & \frac{\Omega}{2} \\ \frac{\Omega}{2} & \delta\omega + ig_{FP} & 0 & 0 & \cdots & 0 \\ \frac{\Omega}{2} & 0 & 2\delta\omega + ig_{FP} & 0 & \cdots & 0 \\ \frac{\Omega}{2} & 0 & 0 & 3\delta\omega + ig_{FP} & \cdots & 0 \\ \vdots & \vdots & \vdots & \vdots & \ddots & \vdots \\ \frac{\Omega}{2} & 0 & 0 & 0 & \cdots & m\omega_{FSR} + \delta\omega + ig_{FP} \end{bmatrix}, \quad (3)$$

where $\delta\omega$ represent the detuning of the FP modes to the QWS mode, and index $m$ denotes the FP modes index spaced by $m \times \omega_{FSR}$. The terms $g_{QWS}$ and $g_{FP}$ denote the net gain of the QWS defect mode and FP modes, respectively. Here we neglect the waveguide losses and consider a cold cavity, where the net gain is given by the mirror losses. This approximation is justified by the fact that when driving the target DFB section, the large value of the difference in the mirror losses is the determinant factor in the value of $\Delta g_{net}$ in our system. We assume the same coupling between the QWS DFB mode and the FP modes. The temperature tuning of the DFB modes frequencies can be approximated by:

$$\omega_{DFB} = 2\pi \frac{c}{2n_{eff}^* \Lambda},$$

where $\Lambda$ and $L$ denote the grating periodicity and total device length, respectively. The temperature tuning is dominated by the temperature tuning of the refractive index, and the temperature tuning of $\Lambda$ and $L$ are considered usually as second order effect[15]. Whereas the angular frequencies of DFB modes depend on the effective refractive index over the length of DFB section ($n_{eff}^*$), FP modes probe the temperature tuning of the effective refractive index of the FP cavity. This explains the different tuning dependence of these two set of modes as function of the injected current. This difference in the temperature tuning rate, when one of the device sections is electrically driven, causes detuning of DFB modes with respect to FP modes as function of current, as shown schematically in Fig. 1(d).

The real and complex part of eigenvalues for this Hamiltonian matrix is shown in 1(e-g), for a given value of $\Delta g$ and varying $\Omega$. For small value of $\frac{\Omega}{\Delta g}$, system operates in weak coupling regime as shown in Fig.1(e); The modes cross in real part and show anti-crossing in the imaginary part of the eigenvalues. On the other extreme, i.e. large value of $\frac{\Omega}{\Delta g}$, the system's eigenvalues undergo level repulsion in the real part, whereas the imaginary parts cross (see Fig.1(g)). This case is similar to the usual strong coupling regime observed in Hermitian system where lossless optical resonators interact[16]. In between these two extremes, the so-called exceptional point singularity happens, where the eigenmodes coalesce in both real and imaginary parts[8,17] (see Fig.1(f)). The coupling strength in the presented devices can be changed either by changing the design parameters, such as modifying the physical distance of the QWS defect from the FP cavity, or by electrically by changing the driving segment of QCL.

In the following we present results of dual section of DFB QCLs, operating at different regime of coupling. All presented QCL lasers were implemented with an inverted buried hetero-structure process[18]. The active region was grown on InP and based on a heterogeneous quantum cascade stack of two bound-to-continuum single wavelength-active regions.

The first consisted of 17 $In_{0.66}Ga_{0.34}As$ and $In_{0.335}Al_{0.665}As$ quantum cascade (QC) periods optimized for emission at 4.3 µm, followed by 18 QC periods optimized for emission at 4.6 µm. The active region was capped with a lattice-matched InGaAs spacer layer, which was used for the definition of the gratings. Two DFB gratings with different grating periodicity and the length of 2 or 2.5 mm were etched in a wet etch process into this spacer layer. Definition of the laser ridge was also done through a wet etch process the leaded to ridge widths varying between 4 to 6 µm. Electrical contacts were deposited to individually provide current to either one of the gratings, effectively forming a front and rear laser section. In order to reduce cross-leakage between the two devices, the thickness of the InP cladding on top of the active waveguide was reduced to 2 µm in between the two devices by means of a shallow wet etch.

We used a one-dimensional transfer matrix method (TMM) simulation to look at the threshold of the cavity modes, when only one section of device was electrically driven. The temperature profile to estimate the tuning was adapted from COMSOL finite element simulations[13]. We only considered the mirror losses in these TMM simulations and ignored the waveguide losses and gain in these TMM simulations. The TMM calculation was performed on the grating structure with temperature profile extracted from a 3D finite element simulation (COMSOL Multiphysics). In all TMM simulations only the target DFB section was pumped, unless stated otherwise. The temperature profile was extended however to the passive section too. The position dependence of the temperature profile T(x), was transformed the profile of the effective refractive index ($n_{eff}$) as function of position and temperature:

$$n_{eff}(\Delta T, x) = n_{eff}(T_0) \times (\beta \times \Delta T(x) + 1)$$

, with the thermal tuning coefficient $\beta = 7.1 \times 10^5$ $K^{-1}$, and the effective refractive index at threshold $n_{eff}(T0) = 3.165$ at 4.3 $\mu m$ wavelength, which was extracted from the laser spectrum at threshold. The grating was terminated on both sides with refractive index of 1 for air, except for the study of AR coating in the supplmenetary materials[19,20].

The laser characterizations were performed using Laboratory Laser Housing (LLH) laser boxes (Alpes Lasers SA, Switzerland) that offers temperature control with Peltier elements and dry atmosphere with $N_2$ purging. The lasers were driven using a direct-current source (QCL 2000, Wavelength Electronics, USA). LIV curves are recorded using a calibrated photodiode. Spectral measurements are acquired by Fourier transform infrared spectroscopy (FTIR), with a maximum spectral resolution of 0.075 cm$^{-1}$ (Vertex 80, Bruker Corp., USA), with the signal acquired by a DTGS detector (for measurements above threshold) or a MCT detector (for sub-threshold measurements).

TABLE I: List of devices in the manuscript.

| Device | total Length (mm) | relative QWS position | $\frac{\Delta n}{n}$ (%) | Figures |
|---|---|---|---|---|
| Device 1 | 4 | middle | 0.5 | Fig. 2 |
| Device 2 | 5 | shifted towards facets | 0.6 | Fig. 3, 4 |

Figure 2 shows the results for a 4 mm long DFB QCL (Device 1 in table I) with QWS located in the middle of the front and back sections. The results of TMM simulation for the intensity profile of the QWS DFB mode and its closest FP modes on the optical spectrum, shows a relatively good overlap of the modes (2 (a)). A refractive index contrast $\frac{\Delta n}{n} \approx$ 0.5 %, extracted from the measured width of the DFB stop-band, was used in these simulations. The results of the sub-threshold spectral map of the laser (Fig. 2 (b)) confirms the expected onset of anti-crossing between the frequency of the main QWS mode and detuned FP modes. One can also note the clear difference in the temperature tuning of these two set of modes. The bandgap of DFB was measured to be 2.2 cm$^{-1}$, and band-edge DFB modes can be identified at the edge of the gap, also repelling the FP modes. This strong coupling dynamics manifest themselves on the behavior of the laser when operated at current densities above the lasing threshold, as shown by the optical spectral map of this laser is shown in Fig. 2(c). Despite the single mode emission of the laser over the whole dynamic range of the laser above threshold, the laser switches on and off multiple times and each time the laser emission frequency experience a jump, with a value consistent with the observed anti-crossing gap in sub-threshold experiments. In order to highlight this different behavior of the laser, the integrated intensity of the lasing mode is shown by solid red curve in the same graph.

In a 5 mm long dual-section DFB QCL (Device 2 in table I), with QWS positioned at $x=0.4L_B$, the overlap of the modes was found to be dramatically different (see Fig.3 (a)).

The latter observation was further confirmed by the absence of anti-crossing between the modes, when the device was measured at sub-threshold current densities (Fig.3 (b)). The strong interaction of the QWS mode with its surrounding FP modes, clearly visible on the intensity modulations of the QWS modes in this case, distinguishes the state of the system from weak coupling regime. In order to highlight this difference, the results of the TMM simulation for this device is shown in Fig.3 (c). Whereas in weak coupling case, a weak modulation of the QWS mode's threshold is expected, in this case the QWS mode's threshold hop to the value of the threshold for FP modes at crossing points. It worth noting that the higher value of the grating contrast $\Delta n/n \approx 0.6\%$ found in this case, is also supporting the more localization of the QWS mode profile and therefore reducing of the mode overlaps compared to the device presented in 2. The above-threshold spectral map characteristics of this device shows again multiple switching on/off above threshold, however with absence of any splitting energy gaps between each off point (see Fig. 3 (d)). The integrated intensity of the lasing mode is shown by solid red curve in the same graph. The latter highlights the impact of the EP on dynamics of the laser.

Although, we used the device design parameters such as the position of the QWS defect in the grating or the length of the device in order to probe different coupling regimes, these dual section devices provide more degrees of freedom when tuning of the coupling strengths is needed. For instance, the electrical pumping scheme in the dual section devices provides a powerful tuning knob to change the dynamics. In the same device shown in Fig. 3, by swapping the pumped section and passive section, the EP degeneracy is lifted and system moves towards strong coupling. As shown in 4(a), in this case the temperature profile tail disturb a considerable length of the DFB section in DFB section (passive section in this case, as marked by blue color). As a result of this distorion, the reflector r$_2$ of the FP cavity is shifted inside the DFB grating. Therefore, the intensity profile of the FP modes extend to towards the DFB section that contains the QWS defect at the target wavelength (Fig. 4 (b)). The simulated spectral map by TMM (Fig. 4 (c)), shows transition from EP regime at small $\Delta T$s to the strong coupling regime and level repulsion at relatively large values of $\Delta T$s. This was confirmed experimentally by the sub-threshold spectral map in Fig.4(d). This tuning feasibility is of crucial importance, as it suggest this device to be used for input-output applications.

The design of the dual section DFB QCLs provides a flexible and robust parity-time (PT) symmetric scheme in order to study dynamics of non-Hermitian systems. The laser dynamics was shown to be strongly modified by the strength of the interaction of modes, where various optical power modulation and lasing wavelength combinations can be realized. The simplicity of theses designs in operation as compared to coupled ring cavities and possibility to tune electrically the strength of interactions make these devices fascinating for exploring the non-Hermitian photonic coupled systems.

**ACKNOWLEDGMENTS**

We acknowledge financial support from the H2020 European Research Council Consolidator Grant (No. 724344) (CHIC), and Nano-Tera.ch foundation under project "IrSens II".

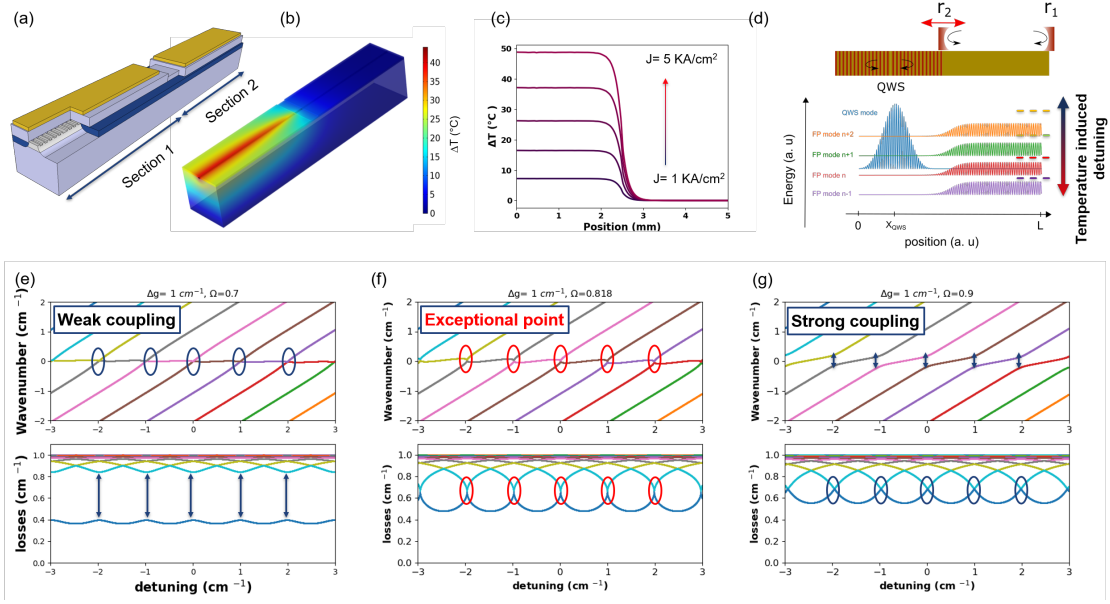

FIG. 1: **The concept of the coupling in dual section DFB QCLs.** (a) Schematic of a dual section DFB QCL with QWS defect in each section. The front and back sections are designed to operate at two different wavelengths, i.e., 4.3 and 4.6 $\mu$m. The etched top cladding between the two sections and the isolated top contacts of the two sections allows driving the two sections independently with a sufficient electrical isolation. (b) Finite element simulation of the thermal distribution along the laser performed by COMSOL multiphysics, when driving one section of device. The color code in the 3D view of device represents the temperature difference to the substrate. (c) The line cut of the temperature difference along the laser ridge for different injection current densities in the laser. (d) Top: The schematic showing the two cavities: the DFB grating cavity with a QWS defect on left side and the FP cavity made by the laser facet reflector ($r_1$) and reflection from the DFB grating ($r_2$) on the other side. Applying different bias current, or biasing different segments of QCL, can change the temperature profile and refractive index profile along the laser. Distortion of the grating refractive index shift the position of the reflector $r_2$. Bottom: The schematic of the field intensity for the QWS mode and the closest FP modes to that, that fall in the stop-band of the grating spectrum. The intensity plots are shifted vertically to show their energy difference. The FP modes, are localized in the passive section of device, whereas the QWS is mainly localized in around the QWS defect position. Different temperature dependent tuning of the FP modes and the DFB modes, lead to detuning of these two set of modes. (e-g) Illustration of the real part (frequency) and imaginary part (losses) of the eigenmodes of the Hamiltonian (3), consisting of a single DFB mode coupled to multiple FP modes. From left to right, depending on the value of $\frac{\Delta g}{\Omega}$, systems undergoes weak coupling regime (mode repulsion in imaginary part), Exceptional point (coalescence of the modes in both real and imaginary part), and strong coupling (repulsion of the modes in real part of eigenvalues, similar to strong coupling in Hermitian systems).

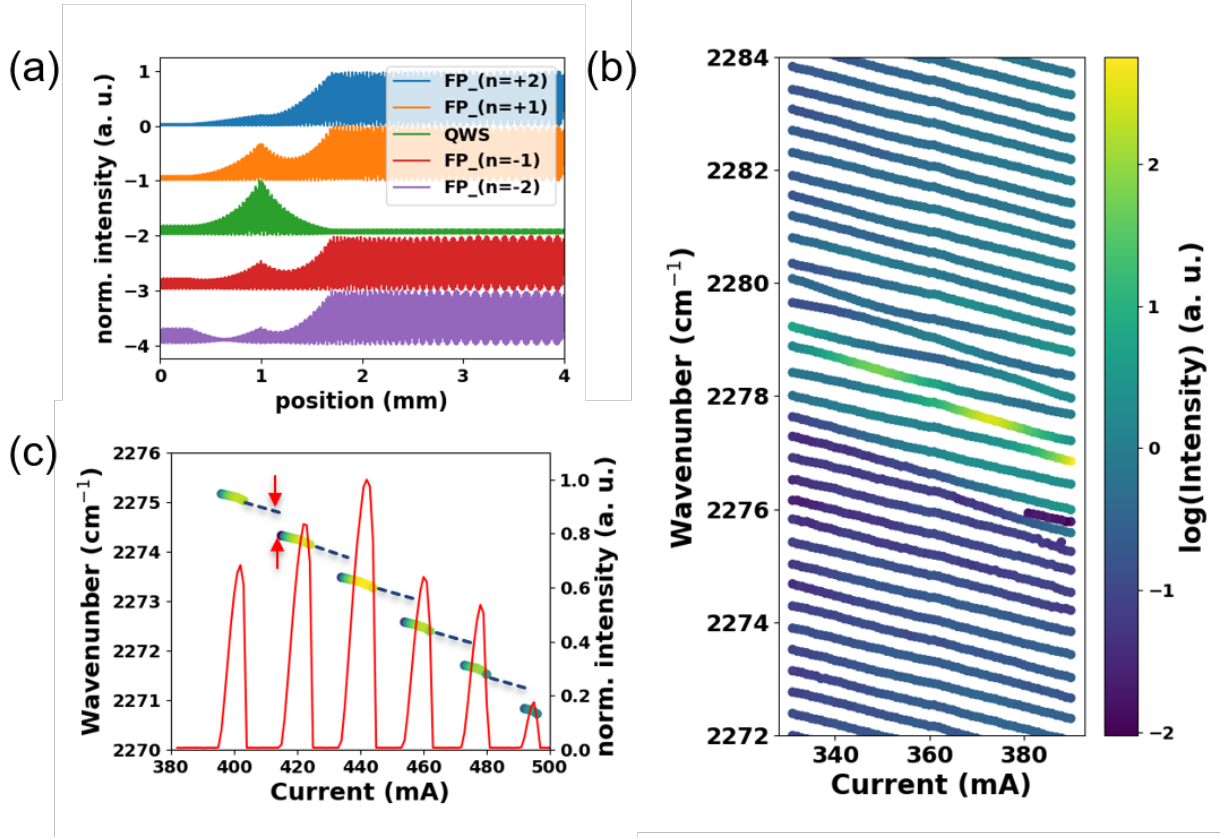

FIG. 2: **Strong coupling of FP modes to the QWS DFB modes.** (a) The intensity profile of the QWS mode and its closest FP modes extracted from TMM simulations for a 4 mm long dual section DFB QCL (Device 1), where the QWS on is positioned at the middle of the driven section and the grating refractive contrast was estimated to be $\frac{\Delta n}{n} \approx 0.5\%$ from the width of the DFB stop-band. The position of the QWS defect in the grating and length of the device allows relatively large overlap of the mode profiles. (b) The measured sub-threshold spectral map of the device. The QWS mode is distinguished by the highest intensity mode in the middle of photonic gap, and anti-crosses with FP modes. (c) Spectral map of the device at driving currents above threshold. The QWS lasing mode turn on and off multiple times, and there is a clear energy gap between each turning off point (marked by dashed lines) and the consequent turning on point. The integrated optical intensity of the QWS mode is plotted in red.

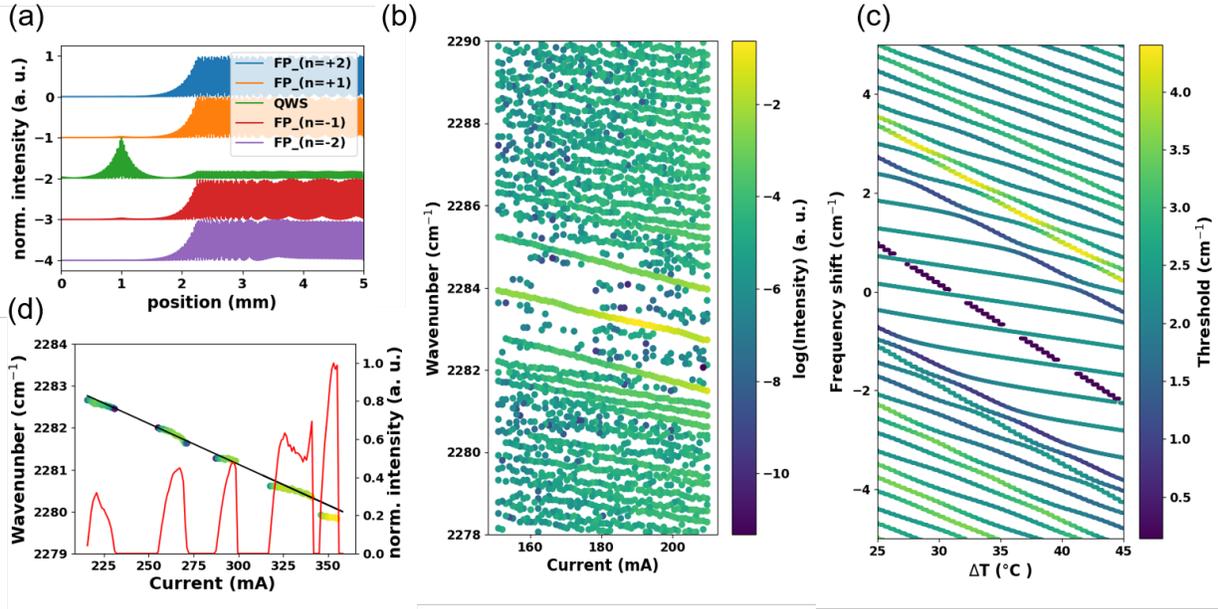

FIG. 3: **Multiple occurrence of EP over the dynamic range of DFB QCL.** (a) The intensity profile of the QWS mode and the closest FP modes to that as function of position along the device extracted from TMM simulations, for a 5mm long dual section DFB QCL (Device 2), where the QWS on is positioned at $0.4L_B$. The grating refractive contrast was estimated to be $\frac{\Delta n}{n} \approx 0.6\%$ from the width of the DFB stop-band.(b) Measured spectral map of the device at sub-threshold currents. The map is obtained when driving the DFB section corresponding to the observation frequency. The absence of the mode repulsion and still strong modulation of the intensity of the QWS mode are the remarkable differences compared to strong/weak coupling regime. (c) TMM simulation results for the grating structure of the device, including the temperature profile along the device waveguide as function of the injected current. The QWS mode crosses in both real (frequency) and imaginary part (losses) when the FP modes tuned to be in resonance with the QWS DFB mode. (d) The measured spectral map of the device when driven above threshold. One distinct feature here is the absence of the frequency gaps between the sequence of off/on points, consistent with the absence of anti-crossing in the sub-threshold spectral map. The integrated intensity of the QWS mode is shown by the red curve, distinguish the dynamics in this case from weak coupling regime, where weak modulations of the output power is expected.

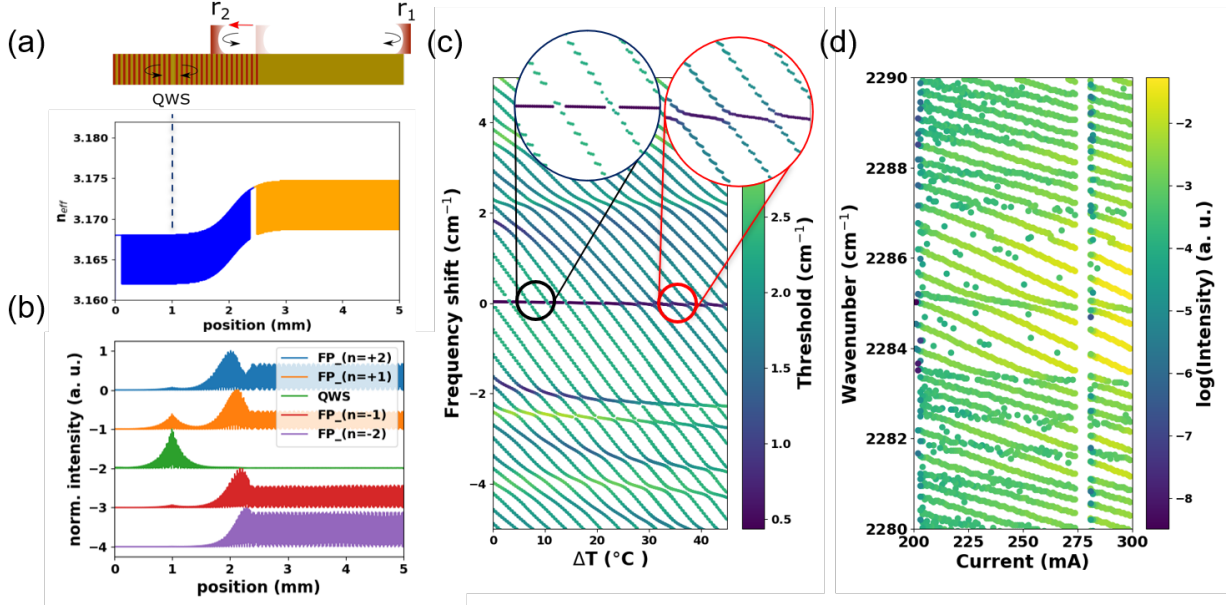

FIG. 4: **Lifting EP degeneracy by different electrical pumping schemes.** (a) The effective refractive index profile along the waveguide of laser, where the QWS defect location is marked by dashed line at x=0.4×$L_B$. The device is driven eclectically in the counterpart section (the orange color) of the one with DFB designed at the observation wavelength. The inset shows schematically the impact of the temperature profile induced by changing the electrically pumped section, as a shift of the reflection point for the FP modes inside the DFB grating ($r_2$). (b) Intensity profile of the the QWS mode and the four closest FP neighbor modes as function of position along the laser waveguide, extracted for $\Delta T$= 35 °C from TMM simulations. As a results of the extension of the heating profile towards the target DFB section, the intensity profile of FP modes penetrate more in the DFB section. (c) The simulated spectral map at sub-threshold current densities. As a consequence of the increased overlap between the FP modes and the QWS mode intensity, the coupling strength $\Omega$ is increased by increasing the temperature difference between the two section $\Delta T$, and the FP modes starts to repel the QWS mode. The insets highlight the onset of coalesce of modes and anti-crossing of modes in real parts for different values of $\Delta T$. (d) The measured sub-threshold spectral map of the device, where EP degeneracy is clearly lifted at the applied current range to the device.

# Supplementary Information: Exceptional Point Singularities in Multi-Section DFB Lasers

Mehran Shahmohammadi,[1, a)] Martin J. Süess,[1] Romain Peretti,[1] Filippos Kapsalidis,[1] Andres Forrer,[1] Mattias Beck,[1] and Jérôme Faist[1]

*Institute for Quantum Electronics, ETH Zürich, 8093 Zürich, Switzerland*

(Dated: 7 March 2022)

---

[a)]correspounding author: shahmohammadimehran@gmail.com

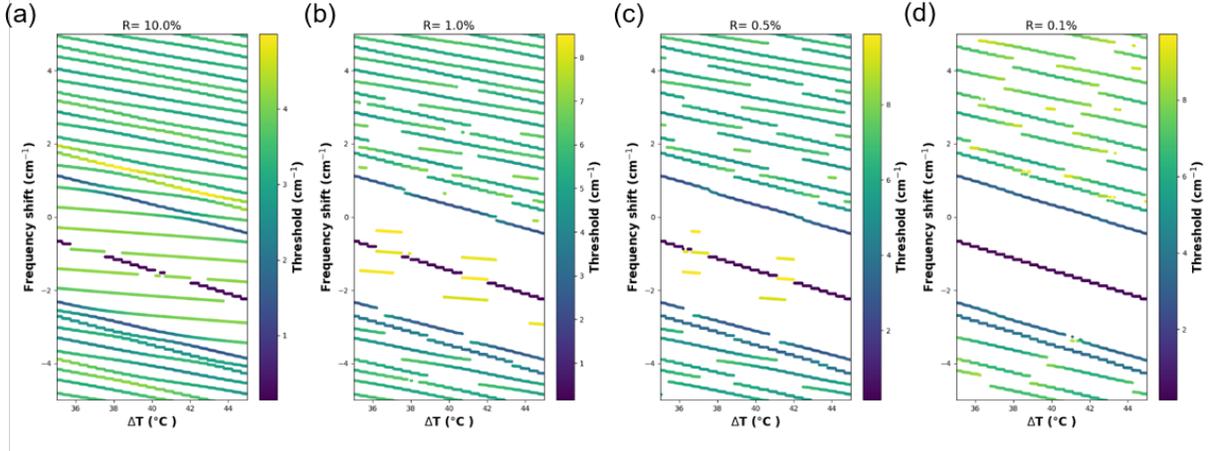

FIG. S1. The effect of AR coating application on both facets, for a 5 mm long DFB QCL, with QWS positioned at $0.4 \times L_B$ and $\frac{\Delta n}{n} \approx 0.6\%$, i.e. the same design parameters as Device 2. A simple $\lambda/4$ anti-reflective coating with refractive index of $\sqrt{n_0 n_s}$ was applied in the simulations. The coating thickness was changed in order to have facet reflectivity of 10%, 1%, 0.5%, and 0.1% at the wavelength of QWS DFB mode, in (a-d) respectively. These simulations show the robustness of the EP in this device design, where only for coatings with reflectivity below 0.5%, the EP degeneracy is lifted.